\documentclass[structabstract]{aa}  
\usepackage{graphicx}
\usepackage{txfonts}
\usepackage{graphicx}
\usepackage{natbib}
\usepackage{pifont}
\begin{document}
   \title{Formation of chondrules in radiative shock waves}
   \subtitle{I. First results, spherical dust particles, stationary shocks}

   \author{H. Joham\inst{1}
          \and
          E. A. Dorfi\inst{2}
		  }

   \institute{Institute for Astronomy, University of Vienna,
              T\"urkenschanzstra{\ss}e 17, A-1180 Wien\\
              \email{a9806222@unet.univie.ac.at}
         \and
             Institute for Astronomy, University of Vienna,
              T\"urkenschanzstra{\ss}e 17, A-1180 Wien\\
             \email{ernst.dorfi@univie.ac.at}
             }

   \date{Received 2012; Accepted 2012}

% \abstract{}{}{}{}{} 
% 5 {} token are mandatory
 
  \abstract
{The formation of chondrules in the protoplanetary nebulae causes many questions concerning the formation process, the source of energy for melting the rims, and the composition of the origin material.}
{The aim of this work is to explore the heating of the chondrule in a single precursor as is typical for radiation hydrodynamical shock waves. 
We take into account the gas-particle friction for the duration of the shock transition and calculate the heat conduction into the chondrules. 
These processes are located in the protoplanetary nebulae at a region around $2.5\,\rm{AU}$, which is considered to be the most likely place of chondrule formation.
The present models are a first step towards computing radiative shock waves occurring in a particle-rich environment.}
{We calculated the shock waves using one-dimensional, time-independent equations of radiation hydrodynamics (RHD) involving realistic gas- and dust opacities and gas-particle friction. 
The evolution of spherical chondrules was followed by solving the heat conduction equation on an adaptive grid.} 
{The results for the shock-heating event are consistent with the cosmochemical constraints of chondrule properties. 
The calculations yield a relative narrow range for density or temperature to meet the requested heating rates of $R>10^4\,\rm{K\,h^{-1}}$ as extracted from cosmochemical constraints. 
Molecular gas, opacities with dust, and a protoplanetary nebula with accretion are necessary requirements for a fast heating process.
The thermal structure in the far post-shock region is not fully consistent with experimental constraints on chondrule formation since the models do not include additional molecular cooling processes.}
   {}


   \maketitle
%
%_______________________________________________________________________________________________

\section{Introduction}
%_______________________________________________________________________________________________

Chondrules (Greek \emph{Chondros}, grain) or chondren are silicate globules with sizes of $0.2$ to $2\,\rm{mm}$ and consist mainly of olivine and pyroxene. 
They make up $20$ to $90\,\rm{\%}$ of the meteorite volume, which is why a whole group of meteorites is called chondrites. 
The mineral structure of the chondrules together with their chemical composition tell a lot about their formation process. 
Based on their observed consistence and structure the heating has to occur very quickly at a rate of $>10^4\,\rm{K\,h^{-1}}$, which strongly favours melting of the origin material. 
A subsequent rapid cooling within a few hours is also required to produce the observed mineralogical features (see, e.g.~\citet{heizmod} for a detail discussion of these basic properties). 
The analysis of various chondrules proves that multiple heating events have been present to generate their final structure. 
Based on these investigations the origin of the chondrules lies within regions of the early protoplanetary nebulae and their age is considered to be between 1 and 4 million years after the formation of the solar system \citep[e.g.][]{amelin,kitetal}.

There exists a large number of models for chondrule formation. 
All of them contain more or less plausible assumptions. 
Owing to the aforementioned structural characteristics of the chondrules, the duration, temperature, and pressure at the heating event have to occur within a spatially localised zone of the protoplanetary nebulae as summarised, e.g.~by \citet{chentst}. 
Shock waves appear to meet most constraints and this scenario is currently the most plausible model, as discussed, e.g.~by \citet{CaL98}, \citet{Desetal}. 
Since the models based on heating in shock waves make fewer assumptions on the physical environment, we will focus on these events, which have been investigated by a large amount of detailed modelling, e.g. \citet{HaH91, HaH93}, \citet{RaI94}, \cite{Hood98}, \citet{DaC02}, \citet{Miura02}, and \citet{MaN05, MaN06}. 
\citet{CaH02} and \citet{Cetal04} developed gas-particle suspension models with significant heating via radiation from the other chondrule-sized particles. 
In the models of \citet{DaC02} the heating process is caused either by frictional heating between gas and dust particles when traversing an adiabatic shock wave or by thermal conduction from the shock-heated gas to the cooler dust particles. 
Desch and Connolly also took significant heating by radiation from other chondrules into account.
Nevertheless, without detailed modelling of the physical processes within the protoplanetary disc the origin of these shock waves remains unanswered. 
In the literature several mechanisms have been suggested to explain the shock waves, i.e.~accretion shocks on the surface of the protoplanetary disc \citep{RaI94}, infall of clumps of gas onto the nebulae \citep{Tanaka98}, bow shocks produced by planetesimals on eccentric orbits \citep{Weidens98}, and X-ray flares from the young Sun \citep{Naka05}. 
Clearly, all these scenarios give only a simplified picture of the numerous interactions between radiation field, and gas, and dust particles. 
In particular the treatment of adiabatic shock waves is mostly made with a simple equation of state and based on the equations of hydrodynamics without calculating a consistent radiative precursor region.

The recent work of \citet{MD2010} is based on a hydrodynamical shock model including a complete molecular line cooling due to ${\rm H_2O}$, a treatment of the radiation field, and improved opacities of the solids. 
These authors found the effects of molecular cooling to be minimal because the combination of high column densities of water, hydrogen recombination/dissociation and radiation from the upstream regions reduce the rapid cooling times of chondrules in the post-shock region.

In this work we investigate the reaction of dust particles to the structure of radiative shock waves. 
Based on full one-dimensional time-independent equations of radiation hydrodynamics (RHD) \citep[e.g.][]{Mihalas}, we follow the evolution of a spherical dust particle that is swept through these shock transitions, which are parameterised by the Mach number. 
In this first study we include the effects of gas-particle drift but the chondrules are still assumed to preserve their spherical shape and are not directly coupled through their radiative properties to the equations of RHD. 
Their temperature structure is obtained by solving a heat conduction equation allowing a thermal expansion of the dust particles. 
Thanks to the full system of RHD, the overall structure of the radiative shock waves is calculated correctly together with the radiative precursor that heats the dust particles before they enter the shock jump.
All models presented are located in a protoplanetary nebula at a typical distance of $2.5\,\rm{AU}$, the most likely  origin of chondrules \citep{heizmod}. 

The next section deals with the basic equations describing the interaction of dust particles, gas, and radiation in one-dimensional plane geometry together with the numerical method. 
In sect.~3 we present the results of several model calculations followed by a discussion of the possible formation process of chondrules in radiating shock waves.

%_______________________________________________________________________________________________

\section{Basic equations}
%_______________________________________________________________________________________________

\subsection{Equations of radiation hydrodynamics (RHD)}
%_______________________________________________________________________________________________

The equations of radiation hydrodynamic (RHD) treat the interplay of matter and radiation by including the momentum and energy exchange between the two components. 
A detailed derivation of these equations can be found in \citet{Landau} and \citet{Mihalas}. 
To investigate the physical behaviour of dust particles we restricted the problem to time-independent planar shock waves in a protoplanetary nebula in this exploratory study. 
We can neglect self-gravity of the gas, and the resulting system of ODEs can be solved without the need of artificial viscosity to broaden shock waves. 
We furthermore assumed the Eddington-approximation as closure condition for the radiation field as well as an ideal molecular gas with a constant adiabatic index of $\gamma=7/5$ for the equation of state.  

The following three equations are the stationary, plane-parallel Euler equations for the gas interacting with the radiation field where $\rho$ denotes the gas density and $u$ the gas velocity. 
We begin with the equation of continuity
\begin{equation}
\label{Kongl}
\frac{d}{dx} \left( \rho u \right) = 0 \:, 
\end{equation}
and the equation of motion with gas pressure $P$ and radiative (Eddington) flux $H$
\begin{equation}
\label{Bewgl}
\frac{d}{dx} \left( \rho u^2 \right) + \frac{dP}{dx} 
        - \frac{4 \pi}{c} \rho \kappa_{\rm R} H = 0 \:,
\end{equation}
where $\kappa_{\rm R}$ is the so-called Rosseland-mean of the opacity, $c$ is the light speed. 
The third equation is the gas energy equation written as
\begin{equation}
\label{Enegl}
\frac{d}{dx} \left( \frac{P u}{\gamma - 1} \right) + P  \frac{du}{dx} 
    - 4 \pi \rho \kappa_{\rm P} (J - S) = 0  \:,
\end{equation}
containing the adiabatic index $\gamma$, the Planck-mean opacity $\kappa_{\rm P}$, the Eddington radiation energy density $J$ and the source function $S$.

The next two equations are needed to describe the radiation field by writing down the Eddington moments of the radiation transport equation. 
Starting with the 0$^{th}$-moment we obtain the equation of the radiation energy density $J$
\begin{equation}  \label{e.J}
\frac{1}{c} \frac{d}{dx} \left( J u \right) + \frac{dH}{dx} 
 + \frac{1}{c} K \frac{du}{dx} + \rho \kappa_{\rm P} (J-S) = 0 \:,
\end{equation}
including the Eddington radiation pressure $K$.
The 1$^{st}$-moment leads to the equation for the radiation flux density $H$, 
\begin{equation}  \label{e.H}
\frac{1}{c} \frac{d}{dx} \left( H u \right) + \frac{dK}{dx} + 
\frac{1}{c} H \frac{du}{dx} + \rho \kappa_{\rm R} H = 0  \:.
\end{equation}

This system of ODEs has to be solved for radiative shock waves together with additional closure conditions, and boundary conditions. 
Hence, we need an equation of state and we adopted in this first study an ideal gas
\begin{equation}  \label{e.eos}
P = \frac{{\cal R}}{\mu} \rho T  \:.
\end{equation}
The source function $S$ is given by 
\begin{eqnarray}
\label{quell}
S = \frac{\sigma}{\pi} T^4
\end{eqnarray}
for LTE, which we assumed throughout. 
$T$ is the temperature and $\sigma$ the Stefan-Boltzmann constant. 
The corresponding radiation temperature $T_{\rm rad}$ can be obtained from the radiation energy density by
\begin{eqnarray} \label{e.trad}
J = \frac{\sigma}{\pi} T_{\rm rad}^4  \: . 
\end{eqnarray}
Since the above equations contain three moments, $J$, $K$, and $H$ of the grey specific radiation intensity, we need an additional closure condition for the radiation field. 
Within the limit of isotropic intensity distributions we can use the well-known Eddington approximation
\begin{equation}
 f = \frac{K}{J} = \frac{1}{3} \: ,
\end{equation}
relating the second to the zeroth moment. 
For the radiating shock waves in a dusty environment such as the solar nebula with its high optical depths this approximation is usually valid.

%_______________________________________________________________________________________________

\subsection{Boundary conditions}
%_______________________________________________________________________________________________

According to the previous section our five basic variables are given by the gas density $\rho$, the gas velocity $u$, the gas pressure $P$, the radiation energy density $J$, and the radiation flux $H$. 
Hence, we have to specify five far up-stream values at $x\to - \infty$. 
Denoting this gas density by $\rho_0$ and the gas velocity by $u_0$, the gas pressure $P_0$ also fixes the gas temperature $T_0$ through the equation of state (\ref{e.eos}). 
From the equation of motion (\ref{Bewgl}) we see that only $H_0=0$ leads to consistent boundary conditions, and the equation of radiation energy (\ref{e.J}) requires $J=S$ and together with the source functions we have $T_0 = T_{{\rm rad,}0}$, i.e~the radiation temperature is equal to the gas temperature for the far up-stream region. 
The same conditions hold for the far down-stream region $x\to + \infty$ for stationary problems \citep[e.g.][]{RHBed}. 
As already discussed in the literature, these conditions appear to be not fulfilled in all shock computations used to study the formation of chondrules. 
Because the radiative flux $H$ is zero at both boundaries, a maximum (or minimum) has to occur in between and any physical possible solution has a maximum located at the shock position $x=0$. 
Ignoring for this discussion the difference between the Rosseland- and Planck mean, we introduce the optical depth $\tau $ by
\begin{equation}
  d\tau = \kappa\rho\, dx
\end{equation}
and can transform the system of stationary RHD equations into
\begin{eqnarray}
   \frac{d}{d\tau} \left( \rho u \right) &=& 0 \\
   \frac{d}{d\tau} \left( \rho u^2 + P \right) - \frac{4\pi}{c} H &=& 0 \\
   \frac{d}{d\tau} \left( \frac{P u}{\gamma - 1} \right) + P  \frac{du}{d\tau} 
                 - 4\pi (J - S) &=& 0       \\
   \frac{d}{d\tau} \left( \frac{Ju}{c} + H \right) 
                 + \frac{K}{c}\frac{du}{d\tau} + (J-S) &=& 0 \\
    \frac{d}{d\tau} \left( \frac{H u}{c} + K \right) + \frac{H}{c}\frac{du}{d\tau} + H &=& 0   \:.           
\end{eqnarray}
We see that the optical depth defines through $(\kappa\rho)^{-1}$ a length scale and hence $(\kappa\rho_0 u_0)^{-1}$ yields a typical time scale available for heating of particles in the precursor region. 

The properties of the shock waves are parameterised by the Mach number $M$ in the up-stream region; $M$ is given for an adiabatic sound velocity $a_0$ by
\begin{equation}  \label{e.mach}
  M = \frac{u_0}{a_0} \quad\mbox{with}\quad a_0^2 = \gamma \frac{P_0}{\rho_0} 
                                                  =  \frac{\gamma{\cal R}}{\mu} T_0 \:.
\end{equation}
A detailed analytical treatment of RHD shock waves can be found in \cite{RHBed}. 
We used the inverse compression ratio $\eta= \rho_0/\rho$, and a value of 
\begin{eqnarray}
\label{eta1}
\eta_1 = \frac{2 + \left(\gamma -1 \right) M^2}{(\gamma + 1 ) M^2 } 
\end{eqnarray}
has to be reached in the far down-stream region behind the shock wave. 

A close view on the aforementioned stationary RHD-equations reveals that the model assumes a single shock wave with a precursor for the melting of the chondrules. 
However, the complexity observed in meteorites favours formation regions where many precursors may exist, e.g.~due to the interaction with bow shocks of moving planetesimals and to accretion processes. 
A more definite answer to the evolution of chondrules therefore requires a more elaborate treatment of the various interaction processes between the gas, dust particles and radiation fields within time-dependent flows. 
In particular, in another step it will be necessary to consider also the momentum and energy exchange between gas and particles because this study only takes into account the heating of the dust particles embedded in the gas without additional energy and momentum exchange. 
Hence, the cooling rates in the post shock region can disagree with data coming from furnace experiments \citep[e.g.][]{HuC05}.

%_______________________________________________________________________________________________

\subsection{Gas and dust opacities}
%_______________________________________________________________________________________________

As mentioned before, we neglected the backreaction of the dust particles on the gas, e.g.~the evolution of the dust particle's size distribution function during the shock transition, which also influences the transparency of the gas-dust mixture. 
The opacities adopted in our computations were computed for evolved planetary discs with a chemical composition like our solar system, i.e.~also a solid-to-gas ration of about $0.01$. 
However, the chondrules may be formed in a dust-enriched environment \citep[e.g.][]{CaH02} where this solid-to-gas ratios can be 100 or even 1000 times the solar ratio. 
In these zones the particle densities are much higher than the adopted values and the passage of a shock wave through this particle-gas suspensions requires a more detailed treatment of the opacities \citep{Desetal}. 
Increasing the dust-to-gas ratio increases the total opacity, and therefore all length scales defined through $d\tau = \kappa \rho\,dx$ will become smaller, in particular the time $t_p$ a chondrule needs to pass the radiative precursors (cf.~Fig.~\ref{schT}) is shortened.

Since we use frequency-integrated moments of the radiation field, the frequency-integrated gas and dust opacities that enter RHD equations, in particular the frequency-dependent opacity $\kappa_{\nu}$, have to be integrated to obtain either the Planck-mean $\kappa_{\rm P}$ for the optically thin case or the Rosseland-mean $\kappa_{\rm R}$ in the optical thick case.
This use of opacity tables for a dust-gas mixture can only be a first approximation to a more realistic situation where the dust particles are processed by the interaction with larger bodies.

In particular, we used the Rosseland- and Planck-mean dust opacities for protoplanetary discs from \citet{opacities}, which are based on optical constants for spherical dust with porous aggregate particles and normal silicates. 

%_______________________________________________________________________________________________

\subsection{Gas-particle friction}
%_______________________________________________________________________________________________

\begin{figure}
   \resizebox{\hsize}{!}{\includegraphics[width=\textwidth]{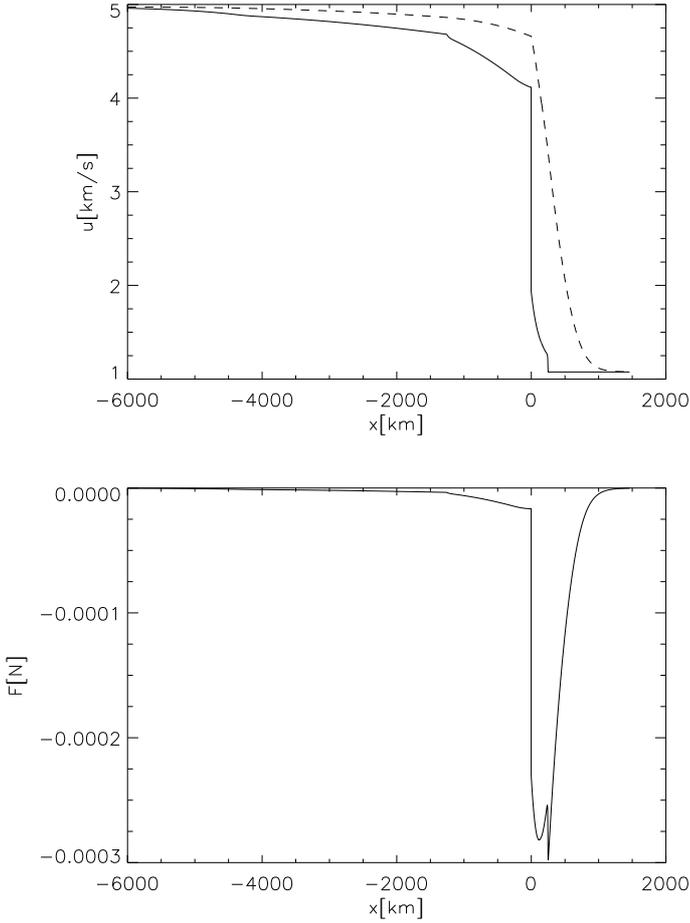}}
    \caption{Gas and particle velocities as a function of the distance from the shock 
    wave in $\rm[km]$ for a Mach number $M=4.1$. 
    Top: Gas velocity and particle velocity (dotted line). 
     Bottom: The drag force on a particle with radius $r_C= 1\, {\rm mm}$.}
\label{Drift}
\end{figure}

When the gas passes the shock wave, the velocity changes in the precursor region, the adiabatic shock, and the subsequent postshock cooling region. 
A particle changes its velocity through collisions with the moving gas and will therefore need some time to relaxate to zero relative velocity. 
The speed of a chondrule $u_C$ will be higher compared to the gas since it maintains its initial up-stream velocities in the absence of friction. 
Consequently, the exposition of dust grains in the hot shock zone is shorter compared to the gas. 
To account for this effect, we solved the following system of equations for a moving chondrule at a position $x_C$
\begin{equation}\label{momentgeschw}
   u_C=\frac{dx_C}{dt}\quad\mbox{or}\quad
   u_C \frac{dt}{dx_C}-1=0  \:.
\end{equation}
The equation of motion for a chondrule can be written as
\begin{equation}
   F_C = m_C \frac{du_C}{dt}\quad\mbox{or}\quad
   F_C \frac{dt}{dx_C} - m_C \frac{du_C}{dx_C} = 0  \:.
\end{equation}
The simple formulae for the frictional force $F_C$  is related to the relative or drift-velocity $u_D$ by 
\begin{equation}
   F_C - \frac{1}{2}\pi r_C^2  \rho C_D u_D^2 = 0,\quad\mbox{with} \quad u_D=u_C - u \:, 
\end{equation}
where $m_C$ is the mass of the particle and $r_C$ its radius. $C_D$ is the drag coefficient given for the numerical calculations according to \citep{drag} 
\begin{equation}
   C_D = C_{D,\rm{diff}} + C_{D,\rm{com}} \:.
\end{equation}
Both terms are calculated as follows
\begin{eqnarray}
   C_{D,\rm{diff}} &=& \frac{2}{3} \frac{\pi^{\frac{1}{2}}(1-\varepsilon)}{S_D} 
      \left( \frac{T_C}{T} \right)^{\frac{1}{2}}  \\
   C_{D,\rm{com}} &=& \frac{2\,sign(S_D)}{S_D^2}  
      \left( 1 + S_D^2 - \frac{1}{(0.17S_D^2 + 0.5\vert S_D\vert + 1)^3} \right)
\end{eqnarray}
with 
\begin{equation}  \label{sd}
   S_D = \frac{u_D}{v_{mp}}, \;\;\;\; v_{mp} = \sqrt{\frac{2 k_B T}{\mu m_H}}  \:.
\end{equation}
$\varepsilon$ defines the type of collision, i.e.~diffusive collisions lead to $\varepsilon = 0$ and specular collisions to $\varepsilon = 1$, $T_C$ denotes the temperature of the particles, $T$ is the temperature of the gas and $v_{mp}$ is the thermal speed of the Maxwellian velocity distribution. 
Figure \ref{Drift} shows the velocity difference between gas and particles and also the force on the particles plotted for a shock wave with Mach number $M=4.1$, a preshock temperature $T_0=300\,\rm{K}$, a preshock density $\rho_0=1.5\!\cdot\!10^{-6}\,\rm{kg\,m^{-3}}$ and an adiabatic index $\gamma=7/5$.
The radius of the particle is set to $r_C=1\,\rm{mm}$ and the density corresponds to $3.2\!\cdot\!10^{3}\rm{kg\,m^{-3}}$ assuming forsterit \rm{Mg$_2$SiO$_4$}.

The stopping length $l_{\rm s}$ of a typical chondrule with radius $r_{C}$ and density $\rho_C$ can be estimated by equating the mass of a dust particle with the gas mass contained in a cylinder of length $l_{\rm s}$ and a cross section corresponding to the dust particle. 
This simple approximation leads roughly to 
\begin{equation}
   l_{\rm s}\simeq r_C \frac{\rho_C}{\rho} \,
\end{equation}
and for the above values (also used in Fig.~\ref{Drift}) we derived $l_{\rm s}\simeq 2000\,$km. 
Hence, we expect relative velocities between gas and chondrules of about $\pm 2000\,{\rm km}$ in the vicinity of the shock waves in agreement with Fig.~\ref{Drift}.

%_______________________________________________________________________________________________

\subsection{Heat conduction in particles \label{waeinst}}
%_______________________________________________________________________________________________

As a first step we studied the heating of a spherical dust particle by heat conduction where the temperature changes because of the passage through the shock at the particle surface. 
In the simplest approximation we assumed a spherical shape of the particle as an initial condition in the upstream region.
The particle is also assumed to be in thermal equilibrium with the surrounding gas, to be located many optical depths away from the radiating shock wave, and to move with the gas, i.e.~no drift velocity $u_D=0$. 

For calculating the particle temperature we solved the one-dimensional conduction equation in spherical geometry. 
Since the gas moves at high velocities and the diameter of the chondrule is small (at maximum a few millimeters), we can assume that the temperature is homogeneous over the spherical surface. 
For further applications we rewrite the heat conduction equation in a volume-integrated form that also allows changes of the radial extension or the geometrical shape for heating (and melting). 
For the particle temperature $T_C$ the spherical heat equation reads
\begin{equation} \label{SWleitgl}
\frac{\partial T_C}{\partial t} = 
   \frac{1}{r^2} \frac{\partial}{\partial r} 
   \left(r^2 D \frac{\partial T_C}{\partial r}\right ) \:,
\end{equation}
where $D$ specifies the heat conduction coefficient. This coefficient can be written as
\begin{equation}
D = \frac{\lambda}{\rho_C c_s} \:,
\end{equation}
$\lambda$ is the heat conduction efficiency, $\rho_C$ is the chondrule density and $c_s$ its specific heat.

For the numerical solution of this PDE we used a standard implicit difference scheme \citep[e.g.][]{Richtmyer}, which leads to a tridiagonal system of equations for the temperature distribution $T_C(r)$ at the different particle radii. 
In this formulation the total radius can change in time due to a thermal expansion of the particle.
The initial condition is specified by a constant temperature within the particle $T(r,0)=T_0(r)$  for $0\leq r\leq r_C$, and later on we assumed an outer boundary condition of $T(r_C,t)=T(x_C)$ as well as $dT_C/dr=0$ at the centre. 
The particle position $x_C$ is determined by integrating Eq.~(\ref{momentgeschw}).

%_______________________________________________________________________________________________

\subsection{Numerical calculations}
%_______________________________________________________________________________________________

For the numerical solution of the RHD system (Eqs.~\ref{Kongl}-\ref{quell}) together with the opacity tables we adopted a standard package SLGA2 \citep{SLGA} for ODEs, written in FORTRAN. 
We sought solutions where $H_0=0$ but $dH_0/d\tau > 0$. 
Together with the remaining up-stream values the equations were integrated towards the shock position and matched by the usual Rankine-Hugoniot conditions to the down-stream values.
A continuous transition of the gas temperature $T$ is not possible but the radiation energy density $J$ and the radiation temperature $T_{rad}$ cross the shock wave without discontinuity.

On top of the RHD shock solutions we followed the path of a dust particle by integrating Eqs.~\ref{momentgeschw}-\ref{sd}.
Numerical experiments show that the type of gas-particle collision is negligible for our calculations and we adopted $\varepsilon=0.5$ throughout. 
During the trajectory the particle temperature $T_C$ is not yet known, but should be in between radiation temperature $T_{\rm rad}$ and gas temperature $T$, and for these computations we considered the gas temperature as the appropriate outer boundary condition for the chondrules, i.e.~$T_C(x,r=r_C)= T(x_C)$.
 
The calculation of the heat conduction in the dust particles was made by numerically solving the tridiagonal equation system, which we obtained from discretisation of the heat conduction equation in spherical symmetry (u.v. \ref{waeinst}). 
We also included the thermal expansion and estimated the amount of radial increase by taking a linear expansion coefficient of $\alpha=9.4\!\cdot\!10^{-6}\,\rm{K^{-1}}$ together with a temperature change of $\Delta T = 1500\,{\rm K}$. 
We obtained an increase of $\Delta r_C = 1.5\!\cdot\! 10^{-2}\,{\rm mm}$ or about 1.5\% for a mm-sized chondrule. 
Clearly, this effect is negligible but we adopted a finite volume discrete version of Eq.~(\ref{SWleitgl}) for subsequent applications, studying also the temporal evolution from fluffy to spherical particles.

%_______________________________________________________________________________________________

\section{Results}
%_______________________________________________________________________________________________

For all RHD shock wave models listed in Table~\ref{t.mod} we considered a spherical dust particle with a radius of $r=1\,\rm{mm}$.  
According to \citet{heizmod}, this size is close to the largest chondrules and we assumed a molecular gas with an adiabatic index of $\gamma=7/5$ throughout. 
According to \citet{HuC05}, the peak temperatures during the forming process of the chondrule are in a range between $T=1700\,-\,2100\,\rm{K}$. 
These gas temperatures destroy all molecules and thereby also change the adiabatic index to a value of $\gamma=5/3$. 
However, to study the effects within the radiative precursor as well as the radiative cooling effects behind the shock waves, we tried to keep the equation of state as simple as possible for the moment. 
The chondrules were assumed to consist of  forsterit (\rm{Mg$_2$SiO$_4$}) with the following particle characteristics. 
The particle density was set to $\rho=3.2\!\cdot\!10^3\rm{kg\,m^{-3}}$, the heat conductivity  $\lambda=1.33\,\rm{W/m\,K}$ and the specific heat $c_s=1.01\,\rm{kJ/kg\,K}$ were taken from the literature \citep[e.g.][]{Kris, Ying}. 
The linear heat expansion coefficient is $\alpha=9.4\!\cdot\!10^{-6}\,\rm{K^{-1}}$.

The protoplanetary disc with accretion has been modelled by \citet{psDisk}, 
where an angular momentum transport coefficient $\alpha=10^{-4}$ leads to a mass accretion 
rate of $\dot{M}=1\!\cdot\!10^{-8}\,\rm{M_{\odot}\,y^{-1}}$. 
Outside of $5\,\rm{AU}$ these discs are gravitationally unstable. 
Taking a typical solar distance of $2.5\,\rm{AU}$ we found a gas density of 
$\rho\sim 10^{-6}\,\rm{kg\,m^{-3}}$ in the mid plane at a temperature $T\sim 300\,\rm{K}$.
According to \citet{heizmod}, the temperature in the chondrule precursor region must not 
exceed $T=650\,\rm{K}$ so that in chondrules the existence of chemical compounds such as \rm{FeS} can be explained. 
Furthermore, the melting time has to be in the range of about $100\,\rm{s}$, which 
requires heating rates of $R>10^4\,\rm{K\,h^{-1}}$ and the subsequent temperature 
decline has to occur over a period of an hour or more \citep{HuC05}.

\begin{figure} 
\resizebox{\hsize}{!}{\includegraphics[width=\textwidth]{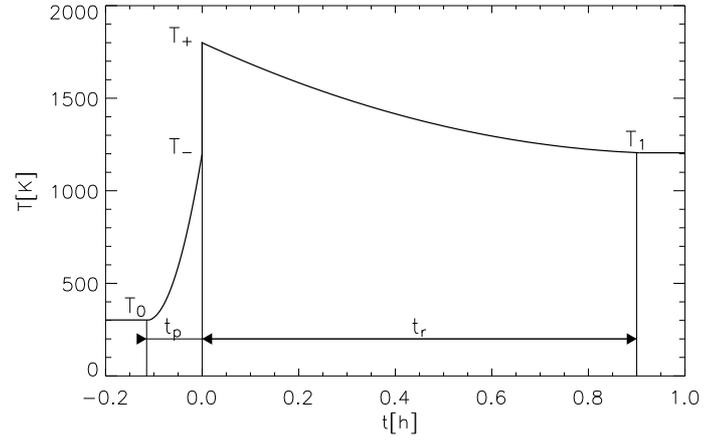}}
\caption{Schematic temperature curve as required by chondrule characteristics 
with preheating wave $T_0=300\,\rm{K}$ up to $T_-=1200\,\rm{K}$ in $t_V=400\,\rm{s}$ ($R>10^4\,\rm{K\,h^{-1}}$). 
The shock front heats the gas to $T_+=1800\,\rm{K}$ and during the subsequent 
relaxation time the temperature drops to $T_1=1200\,\rm{K}$.}
\label{schT}
\end{figure}

\begin{figure}
\resizebox{\hsize}{!}{\includegraphics[width=\textwidth]{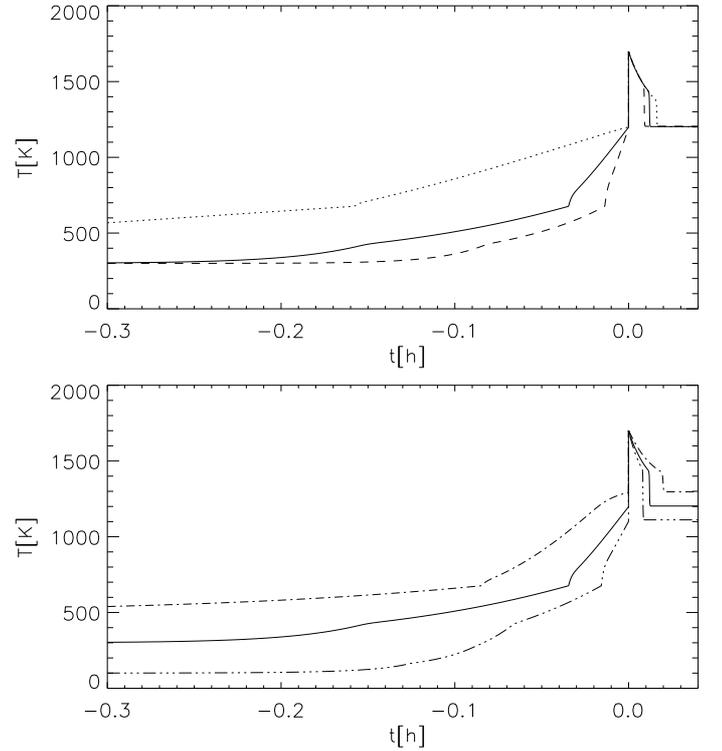}}
\caption{The temporal structure of the temperature within a RHD shock wave reaching a maximum of 
   $T_{peak}=1700\,\rm{K}$ in units of hours relative to the passage through the shock wave. 
   The top panel plots the variations in $\rho_0$, $T_0=300\rm{K}=\rm{const}$; bottom: Variations in 
   $T_0$, $\rho_0=2\!\cdot\!10^{-6}\,\rm{kg\,m^{-3}}=\rm{const}$; solid line: 
    model 1, dotted line: model 2, dashed line: model 3, dash-dot line: model 4, dash-dot-dot line: model 5}
\label{DTmax1}
\end{figure}

\begin{figure}
\resizebox{\hsize}{!}{\includegraphics[width=\textwidth]{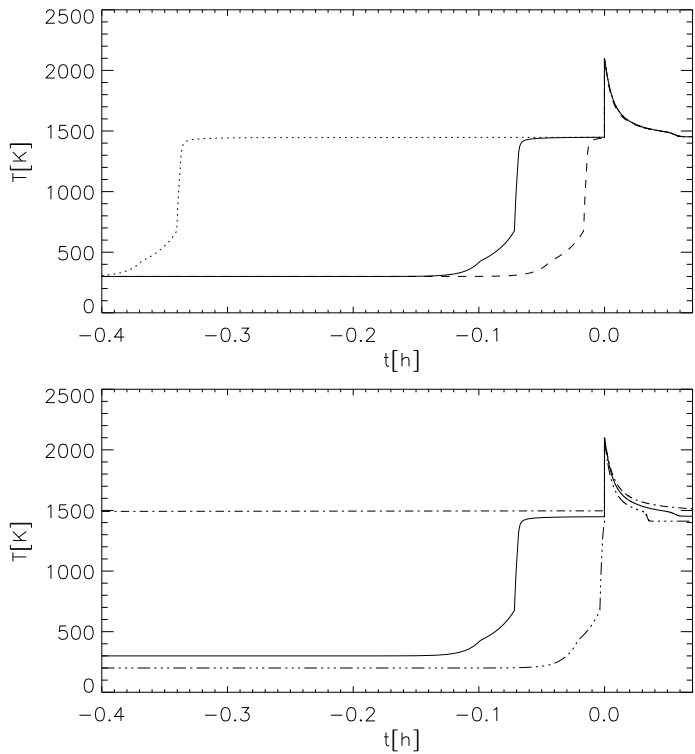}}
\caption{Heating to $T_{peak}=2100\,\rm{K}$ within RHD Shock waves; top: variations in 
  $\rho_0$, $T_0=300\rm{K}=\rm{const}$; bottom: variations in 
   $T_0$, $\rho_0=5.83\!\cdot\!10^{-6}\,\rm{kg\,m^{-3}}=\rm{const}$; 
   solid line: model 6, dotted line: model 7, dashed line: model 8, dash-dot line: model 9, 
   dash-dot-dot line: model 10}
\label{DTmax2}
\end{figure}

In Figure~\ref{schT} a schematic curve illustrates the basic temperature structure as required from the cosmochemical evidence found within chondrules. 
Table \ref{schT} lists several models with a maximum temperature between $T_{peak}=1700\,\rm{K}$ and $T_{peak}=2100\,\rm{K}$ (see figure \ref{DTmax1} and \ref{DTmax2}). 
The initial density $\rho_0$ and the temperature $T_0$ in the up-stream region ($x_p, t_p$) correspond to different Mach numbers $M$ of the shock transitions. 
The adopted numbers vary around values suggested by simple models of the protoplanetary nebulae.

Since the spatial structures within a radiative shock can be approximated by exponential functions, see, e.g.~\citet{RHBed}, we assumed for our analysis of the precursor region~($x_p, t_p$) and the relaxation zone~($x_r, t_r$)
\begin{equation}
   T(x) \simeq T_{-}e^{-|x|/x_{\rm p}}\quad\mbox{and}\quad
   T(x) \simeq T_{+}e^{-|x|/x_{\rm r}} \:,
\end{equation}
which also defines the flow time through these regions by
\begin{equation}
 t_{\rm p} = \int_{-x_{\rm p}}^0 \frac{dx}{u(x)}\quad\mbox{and}\quad
 t_{\rm r} = \int_ 0^{x_{\rm r}} \frac{dx}{u(x)} \:.
\end{equation}
$x$ denotes the distance from the shock wave located at $x=0$.
The quantiles are also given in Table~\ref{t.mod} allowing a more detailed characterisation of the shock structures with their typical length and time scales shaping the dust particles.

\begin{figure}
\resizebox{\hsize}{!}{\includegraphics[width=\textwidth]{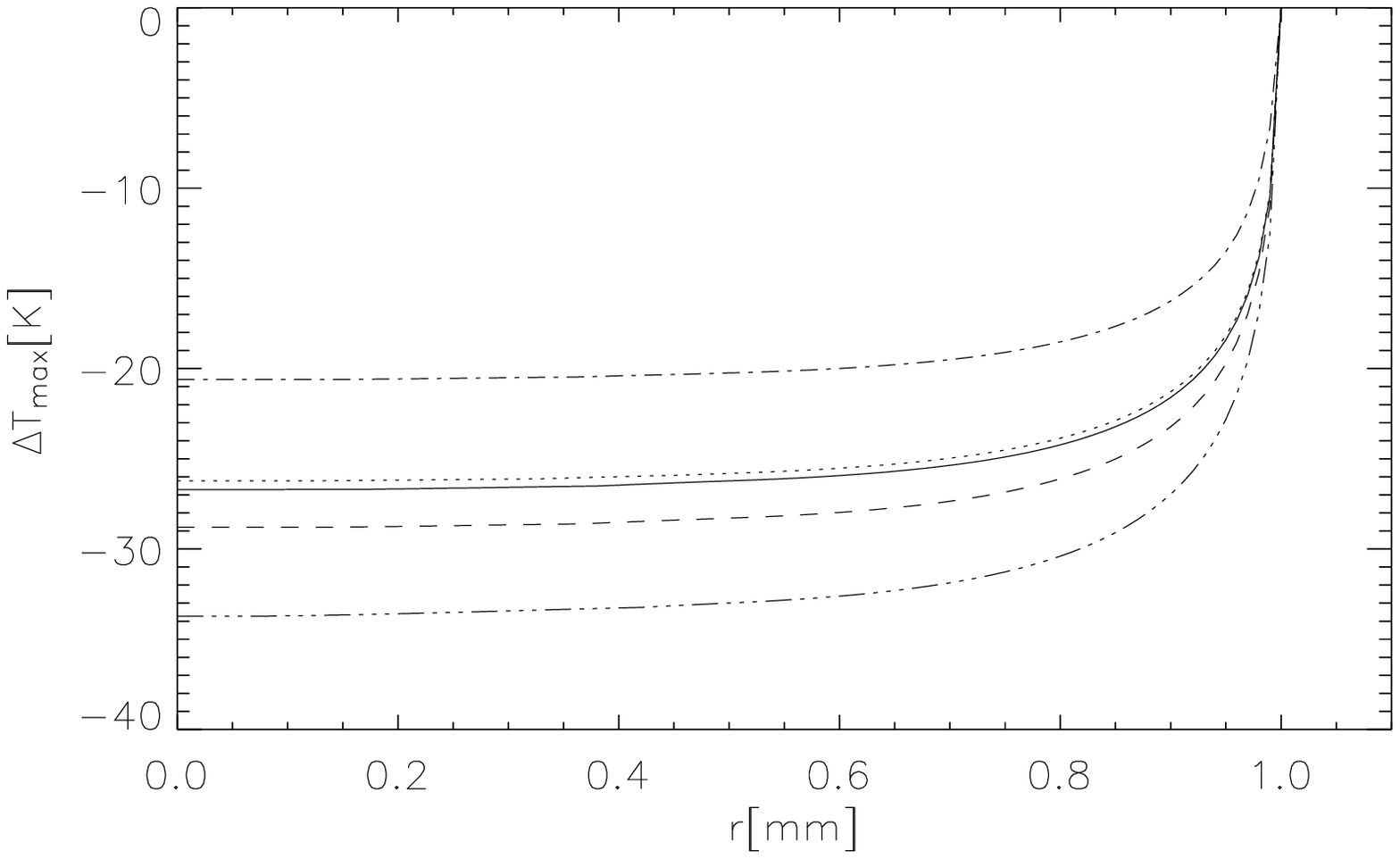}}
\caption{Maximum relative temperature difference within a spherical 
   chondrule as a function of its radius in $[mm]$ for shock waves with $T_{peak}=2100\,\rm{K}$ as 
  summarised in Table~\ref{t.mod}. 
  solid line: model 6, dotted line: model 7, dashed line: model 8, dash-dot line: model 9, 
  dash-dot-dot line: model 10}
\label{Tdiff2}
\end{figure}

The temperature difference between the surface and the center of a chondrule is relative small, as seen from the values of $\Delta T_{max}$. 
Figure \ref{Tdiff2} shows the maximum temperature in a spherical body in dependence of its radius with $T_{peak}=2100\,\rm{K}$ by transition of RHD shock fronts. 

\begin{table}
\caption{Parameter of RHD shock waves with $\gamma =7/5$}
\label{t.mod}
\centering
\begin{tabular}{|l|c|c|c|c|c|}
\hline
\hline
Model	&1&2&3&4&5\\
\hline
$\rho_0\, [10^{-6}{\rm kg\,m^{-3}}]$	& 2.0 & 1.0 & 3.0 & 2.0 & 2.0 \\
\hline
$T_0\,\rm{[K]}$	&\multicolumn{3}{|c|}{300}& 500 & 100 \\
\hline
$T_+\,\rm{[K]}$	&\multicolumn{5}{|c|}{$1700$}\\
\hline
$M$ &  3.97 & 3.97 & 3.98 & 2.92 & 7.23 \\
\hline
\hline
$x_p\,\rm{[km]}$ & $83$ & $3600$ & $370$ & $1100$ & $980$\\
\hline
$x_r\,\rm{[km]}$ & $175$ & $225$ & $132$ & $220$ & $131$\\
\hline
$t_p\,\rm{[s]}$ & $179$ & $780$ & $79$ & $240$ & $199$\\
\hline
$t_r\,\rm{[s]}$ & $43$ & $55$ & $32$ & $60$ & $30$\\
\hline
$R\, [10^4\,{\rm K\,h^{-1}}]$	& $\sim 2.4$ & $\sim 0.6$ & $\sim 5.8 $ & $\sim 0.9$ & $\sim 5.3 $ \\
%\hline
%$t_{R}<1\,\rm{h}$	&\ding{51}&\ding{51}&\ding{51}&\ding{51}&\ding{51}\\
\hline
$\Delta T_{max}\,\rm{[K]}$	&$13.2$&$12.8$&$13.4$&$9.0$&$17.2$\\
\hline
\multicolumn{6}{c}{}\\
\hline
\hline
Model	&6&7&8&9&10\\
\hline
$\rho_0\,[10^{-6}{\rm kg\,m^{-3}}]$	& 5.83 & 5.7 & 6.0 & \multicolumn{2}{|c|}{5.83} \\
\hline
$T_0\,\rm{[K]}$	&\multicolumn{3}{|c|}{300} & 400 & 200 \\
\hline
$T_+\,\rm{[K]}$	&\multicolumn{5}{|c|}{2100}\\
\hline
$M$ & 4.48 & 4.48 & 4.48 & 3.80 & 5.61\\
\hline
\hline
$x_p\,\rm{[km]}$ & $1150$ & $5600$ & $310$ & $76000$ & $106$\\
\hline
$x_r\,\rm{[km]}$ & $122$ & $122$ & $122$ & $125$ & $119$\\
\hline
$t_p\,\rm{[s]}$ & $232$ & $1220$ & $59$ & $17100$ & $20$\\
\hline
$t_r\,\rm{[s]}$ & $36$ & $36$ & $36$ & $38$ & $32$\\
\hline
$R\,[10^4\,{\rm K\,h^{-1}}]$ & $\sim 2.2$ & $\sim 0.4 $ & $\sim 8.1$ & $\sim 0.03$ 
                             & $\sim 30.7$ \\
%\hline
%$t_{R}<1\,\rm{h}$	&\ding{51}&\ding{51}&\ding{51}&\ding{51}&\ding{51}\\
\hline
$\Delta T_{max}\,\rm{[K]}$	& 26.8 & 26.2 & 28.8 & 20.7 & 33.7 \\
\hline
\end{tabular}
\end{table}

%_______________________________________________________________________________________________

\section{Discussion}
%_______________________________________________________________________________________________

For model 1 with initial temperature $T_0=300\,\rm{K}$ and peak temperature  $T_{peak}=1700\,\rm{K}$ an initial density $\rho_0=\,2.0\!\cdot\!10^{-6}\,\rm{kg\,m^{-3}}$ was chosen so that the requirement of a heating rate $R>10^4\,\rm{K\,h^{-1}}$ is fulfilled.
Model~2 shows that at initial density of $\rho_0=1.0\!\cdot\!10^{-6}\,\rm{kg\,m^{-3}}$ the necessary heating rate is not achieved.  
A higher initial density $\rho_0$, like in model~3, leads to significantly higher heating rates. 
Models~4 and 5 show variations of the initial temperature $T_0$, based on model~1. 
A temperature decrease leads to higher heating rates (model 5). 
The necessary heating rates cannot be reached any more by model 4. 
Models~1 to~5 are almost subcritical shock waves, models~6 to 10 with peak temperatures $T_{peak}=2100\,\rm{K}$ constitute supercritical shock waves. 
With density and temperature variations models 6 to 10 react in the same way as models 1 to 5, but the changes have a greater effect on the duration of the preheating wave and thereby on the heating rate. 
The increase of temperature in supercritical shock waves at the beginning of the preheating waves fulfills the necessary heating rates. 
The particles stay for a long time in the preheating wave until the shock occurs.

Investigating the range of possible initial conditions as constrained by models of protoplanetary discs, we found that the gas can be heated up to $T_{peak}=2100\,\rm{K}$ and that the Rosseland- and Planck-mean opacities with dust can meet the basic requirements of shock-based models. 
These RHD shock waves provide the requested heating rates. 
Calculations using a monoatomic gas with $\gamma=5/3$, opacities without dust (Alexander opacities of \citet{opala}) and a disc models without accretion (e.g.~the minimum mass nebulae of \citet{psNebel}) reveal that the particles are propagating within the preheating wave for days to years, which does not match the required heating rates.
Molecular gas and the dust-enriched opacities result in better agreements with the observational facts of chondrule research. 
The models of a protoplanetary disc including accretion \citep{psDisk} have much higher gas densities in the mid plane around $2.5\,\rm{AU}$. 
This gas densities are essential to compute the requested heating rates. 
Summarising we emphasize that the increase of the up-stream density $\rho_0$ together with a decrease of the temperature $T_0$ makes molecules and therefore $\gamma=7/5$ as well as higher opacities possible. 
All these effects reduce the duration of the dust particle in the preheating wave and increase the heating rates.

The calculated temperature differences between surface and core of the chondrules (Fig.~\ref{Tdiff2}) are correct as long as the chondrules remain solid particles. 
But up to now we have not incorporated phase transitions in this chondrule model. 
Clearly, strong shock waves with high Mach numbers $M$ as well as shock waves with smaller preheating waves will produce larger temperature differences but are unlikely to ensure the stringent cosmochemical constraints. 
As depicted in model 10, the preheating wave vanishes completely and the sudden increase of the temperature at the outer boundary yields the largest temperature differences within chondrules. 
For these conditions it is easier to explain partial or shell melting of the particles. 

Immediately after the shock passage we observe a rapid cooling on a time scale of minutes owing to radiative losses determined by the opacities. 
This effect is followed by a longer time scale of several days where non-equilibrium processes further reduce the post shock temperature of the gas and the chondrules. 
The rapid cooling rates of our models of about $10^4$K/hr seem to disagree with furnace experiments, which indicate values of around $100\,$K/hr. 
The reason for this discrepancy is due to different time scales involved during cooling processes in the post shock region. 
The calculations of, e.g.~\citet{Desetal} include the energy and momentum exchange between the chondrules and the gas. 
Therefore the temperature structure in the post shock region is controlled over longer time scales by an additional cooling term $\Lambda(x)$, which includes the radiative cooling by $\rm H_2O$-molecules. 
Consequently, the far down-stream gas temperature can be set to the far up-stream temperature $T_0$. 
This leads to overall cooling rates of $100\,$K/hr, which cannot be obtained within our physical description, which neglects this additional cooling term. 
The boundary condition demands $T_0=T_1$ for the gas temperature as specified, e.g.~in \citet{Desetal}, and the post shock region cools down to the initial temperature by radiative looses into the surrounding protosolar nebula. 
However, this additional cooling occurring on a longer spatial and temporal scale reduces the overall cooling rates to less than $100\,$K/hr. 
From furnace experiments \citep{HuC05} it has become clear that the exact shape of cooling path has no influence on the final cooling rate.

The recent work of \citet{MD2010} reveals how additional physical mechanisms like molecular line cooling due to water influence the cooling history in the post-shock region. 
However, these authors emphasize that the post-shock region is only slightly changed because several effects partly cancel each other out, i.e.~radiative heating from the shock front as well as recombination and dissociation of molecules. 
Summarising these recent results of \citet{MD2010} we note that the improved nebular shock models can meet all cosmochemical constraints of condrule formation. 
The shock speeds necessary to melt chondrules have to be increased, e.g.~from $7\,{\rm km/s}$ to $8\,{\rm km/s}$.

Several models have been developed during the past years to calculate the behaviour of dust particles within shock waves \citep[e.g.][]{DaC02,MaN05}. 
However, none of the proposed models have included the full set of RHD equations with the most recent gas and dust opacities, which are essential for the correct structure of the precursor region. 
This zone is shaped by the radiation transmitted from the shocked down-stream gas and the dust particles are advected through this region towards the shock wave. 
As inferred from the cosmochemical evidence, this region provides a basic heating process and will therefore restrict the possible Mach numbers. 
As seen from the models (cf.~Table~\ref{t.mod}), already moderate Mach numbers around $M\simeq 4$ (for $T=300\,\rm{K}$) meet the constraints of chondrule formation with parameters reasonably within protoplanetary nebula models. 

Opacities changes due to dust-enriched environments caused by fragmentation and collisional debris \citep[e.g.][]{CaH02} may significantly increase solid-to-gas ratios, and values of $100$ or even $1000$ times the solar ratio can be expected. 
Clearly, shock waves propagating in such a densely dust-populated environment requires a more detailed treatment of the opacities \citep{Desetal}, as adopted in these simulations. 
In particular, all time- and length scales controlled by the opacity are then changed.

As mentioned already, the shape and/or the mass of the particles can be modified as they propagate through the shock region. 
First, the partial or total melting could lead to a mass loss by evaporation or, depending of the surface tension, also to a rearrangement of the surface, i.e.~the particles can become more spherical. 
Secondly, drift velocities acting on the surface can result in an erosion and thereby decrease the particles mass. 
Thirdly, the radiative interactions with the dust particles together with the additional heat needed to melt the particle should be included into these models of radiative shock-heated chondrules. 
The dissociation of molecules also demands a more realistic treatment of the equation of state. 
These structural effects of the dust particles will be discussed and included in a forthcoming paper in more detail.

\begin{acknowledgements}
The publication is supported by the Austrian Science Fund (FWF).
\end{acknowledgements}

%_______________________________________________________________________________________________

\bibliographystyle{aa}

%_______________________________________________________________________________________________

\end{document}